\documentclass[aps,pre,reprint,amsmath,amssymb,floatfix,nofootinbib]{revtex4-2}

\usepackage{graphicx}
\usepackage{tikz}
\usepackage{amsthm}
\usepackage{xcolor}
\usepackage{hyperref}
\hypersetup{colorlinks=true, linkcolor=blue, citecolor=blue, urlcolor=blue}

\providecommand{\apj}{ApJ}
\providecommand{\apjl}{ApJL}

\providecommand{\aap}{A\&A}

\providecommand{\mnras}{MNRAS}
\providecommand{\nat}{Nature}
\providecommand{\physrep}{Phys.~Rep.}
\providecommand{\prd}{Phys.~Rev.~D}
\providecommand{\pre}{Phys.~Rev.~E}
\providecommand{\prl}{Phys.~Rev.~Lett.}

\providecommand{\jgr}{J.~Geophys.~Res.}

\newcommand{\Da}{\Delta\alpha}
\newcommand{\Tcal}{\mathcal{T}}
\newtheorem{result}{Result}

\begin{document}

\title{Exact topology of conservative multiplicative cascades:\\
An ultrametric transfer-operator genus}

\author{Cristiano G. Sabiu}
\affiliation{Natural Science Research Institute (NSRI), University of Seoul,
Seoul 02504, Republic of Korea}
\email[Contact author: ]{csabiu@gmail.com}

\date{\today}

\begin{abstract}
The multiplicative cascade is a well-known method for generating random fields that are both genuinely non-Gaussian and scale invariant. Its one-point and scaling statistics, namely the multifractal spectra $\tau(q)$, $D_q$, and $f(\alpha)$, are known exactly; however, its geometric and topological measures are not.  In this work, we show that the topological
structures of a conservative (microcanonical) random permutation cascade can be precisely predicted.
We compute the digital (cubical)
Euler characteristic, or genus, of the level-$j$ excursion set of the
two-dimensional cascade in closed form, using
a transfer operator on the cascade's $b$-ary tree closed by its ultrametric
structure and a conservative without-replacement sibling split.
The result requires no Monte Carlo and matches simulated realizations with a maximum residual of $\sim\!10^{-4}$ in the Euler density, a controlled discretization artifact of the atomic measure. We tie
the genus scaling analytically to the multifractal spectrum, and show that for a
geometric ladder of weights the genus is exactly self-similar under the
multifractal-width dial. The construction uses the tree-sum cumulant
machinery of Greiner \emph{et al.}\ [Phys.\ Rev.\ E \textbf{58}, 554 (1998)], specialized to
topological functionals; the conservative split carries the microcanonical
fingerprint $\mathrm{Cov}(\ln w_a,\ln w_b)/\mathrm{Var}(\ln w)=-1/(n-1)$ that distinguishes
it from canonical and lognormal cascades. The cascade is the multiplicative
counterpart of the Rayleigh--L\'evy flight and a controlled benchmark for the
morphology of strongly non-Gaussian fields.
\end{abstract}

\maketitle

\section{Introduction}
\label{sec:intro}

The multiplicative cascade~\cite{Mandelbrot1974,FrischParisi1985,
MeneveauSreenivasan1987} is one of two standard methods for constructing random fields that are genuinely non-Gaussian and, at the same time, scale invariant;
the other is the hierarchical (L\'evy) series~\cite{Peebles1980,BernardeauPichon2024}. The conservative cascade redistributes a conserved quantity across a
hierarchy of cells that becomes progressively finer at each step, and the core structure of the resulting multifractal measure
(mass exponent $\tau(q)$, generalized dimension $D_q$, and singularity
spectrum $f(\alpha)$) is precisely determined by a set of cascade weights~\cite{Halsey1986}.
These fields are found wherever intermittent and scale-invariant fluctuations occur: in fully
developed turbulence~\cite{MeneveauSreenivasan1987}, in rain, cloud and
atmospheric fields~\cite{SchertzerLovejoy1987,GuptaWaymire1993,LovejoySchertzer2013},
in financial time series~\cite{Ghashghaie1996}, and in cosmological large-scale
structure~\cite{CastagnoliProvenzale1991,SylosLabini1998,Sabiu2007}. 

The one-point statistics and the scaling statistics of a
cascade are covered in textbooks. However,
\emph{geometric and topological} descriptors are not. The Euler characteristic, or genus, of an
excursion set (more generally, the Minkowski functionals) is a standard, motion-invariant, additive measure of the morphology of a random
field~\cite{MeckeBuchertWagner1994,SchmalzingBuchert1997,OhserMucklich2000},
which captures the phase information that
two-point statistics miss~\cite{ColesChiang2000}. (We use ``genus''
interchangeably with the Euler characteristic; the conventional cosmological
genus differs from $\chi$ only by sign and normalization conventions.)
However, analytical predictions for these
exist only in two domains: in the Gaussian limit, where the Euler density
takes the form of the Tomita/Adler--Taylor function~\cite{Tomita1986,AdlerTaylor2010}, and in the
weakly non-Gaussian regime, where a systematic expansion in low-order cumulants
provides the main correction terms~\cite{Matsubara2003,GayPichonPogosyan2012,Codis2013}.
For \emph{strongly} non-Gaussian fields, there is no corresponding general theory, and
examples that can be solved exactly are rare.

A closely related additive model has recently been solved in controllable
form: the Rayleigh--L\'evy (RL) flight, whose one-point statistics, perimeter,
and critical points have been derived in one and two dimensions via
large-deviation expansion~\cite{Peebles1980,BernardeauPichon2024,Bernardo2026},
with the Euler characteristic so far available only at the \emph{mean-field}
level~\cite{Bernardo2026}. The conservative cascade is the natural multiplicative counterpart to 
this additive testbed: both are hierarchical and non-perturbatively non-Gaussian, and both tend 
to Gaussian under smoothing. What the cascade adds is control over the \emph{shape} of the 
intrinsic non-Gaussianity through its weights (summarized by the singularity-spectrum width 
$\Da$), where the flight's cumulant hierarchy is fixed; what it gives up is symmetry, carrying 
the discrete point and scale symmetries of its grid rather than the flight's homogeneity and isotropy.

\begin{figure*}
\centering
\includegraphics[width=0.95\textwidth]{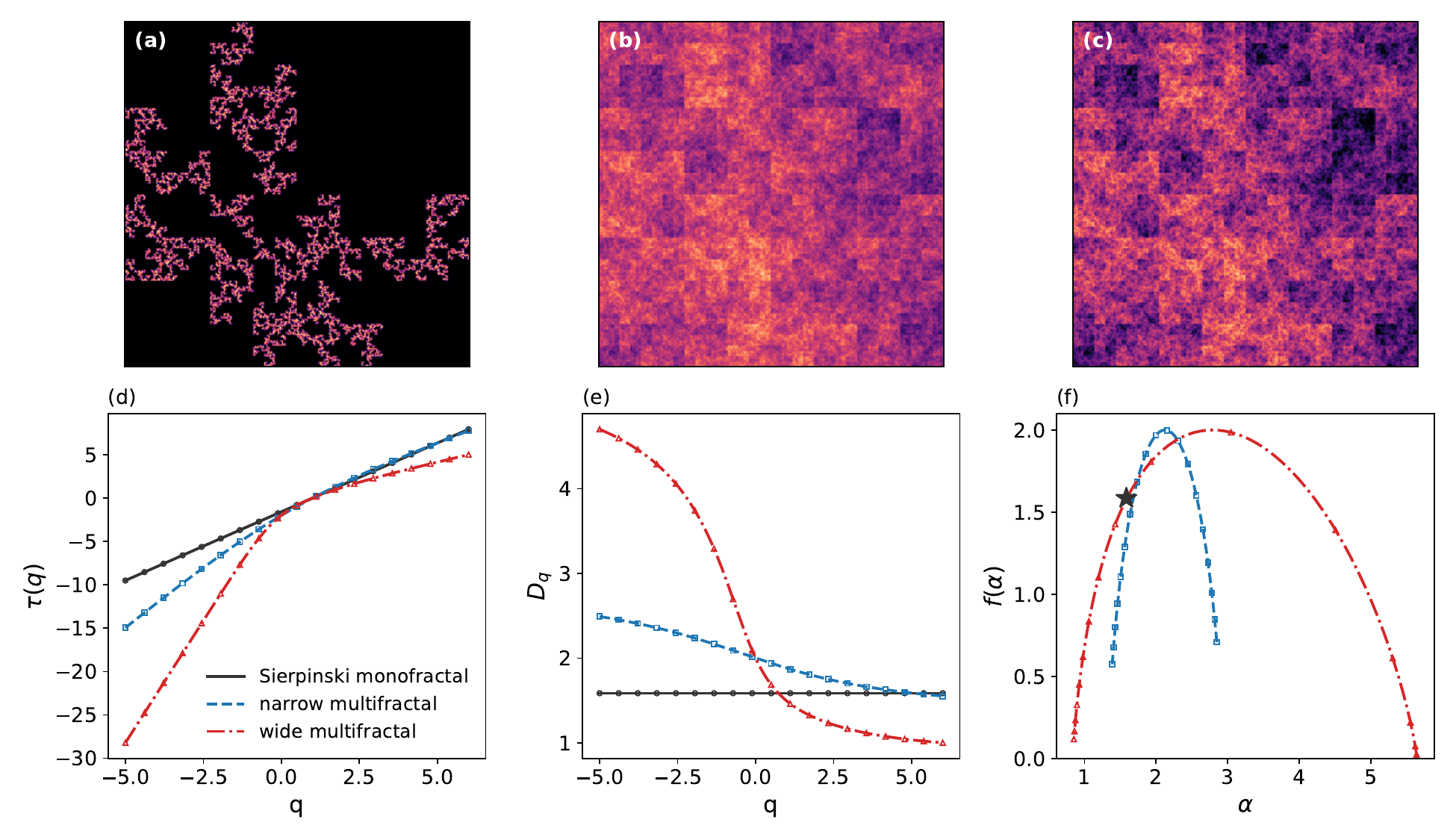}
\caption{Three conservative cascades of increasing singularity width on a shared
random-permutation skeleton: a Sierpinski monofractal
$[\tfrac13,\tfrac13,\tfrac13,0]$ ($\Da=0$), a narrow multifractal $[0.40, 0.28, 0.19, 0.13]$, and a wide
one  $[0.56, 0.28, 0.14, 0.02]$. \emph{(a)--(c):} the $\log_{10}\rho$ density fields ($b=2$, $512^2$),
ranging from a fractal \emph{set} (monofractal) to a strongly intermittent web of
voids, ridges, and peaks (wide). \emph{(d)--(f):} the exact multifractal spectrum
$\tau(q)$, $D_q$ and $f(\alpha)$ of Eqs.~\eqref{eq:tau}--\eqref{eq:falpha}
(lines) versus the Chhabra--Jensen direct measurement (points), exact to machine
precision.}
\label{fig:cascades}
\end{figure*}

In this work we show that the cascade's topology is exactly solvable precisely where
the additive sibling's is not. Our central result is a closed-form expression for
the digital (cubical) Euler density of the two-dimensional conservative cascade measure, obtained
from a transfer operator that walks down the cascade's $b$-ary tree
(Sec.~\ref{sec:genus}).
We validate the prediction against direct simulations and find agreement
to a maximum residual of $\sim\!10^{-4}$ in the Euler density
(Sec.~\ref{sec:validation}); this residual is not statistical or model error
but a controlled discretization artifact of the atomic measure.
The transfer-operator and tree-sum machinery is that of Greiner
\emph{et al.}~\cite{GreinerEggersLipa1998,Greiner1998}, developed for the
correlation cumulants of branching cascades; what is new here is the
conservative without-replacement closure of the joint exceedance
probabilities, the finite-level plaquette alignment combinatorics, and the
closed-form topological functionals they yield, together with the genus
self-similarity under the width dial (Sec.~\ref{sec:scaling}).

\section{The conservative multiplicative cascade}
\label{sec:model}

\subsection{Definition and mass conservation}

We use a conservative (microcanonical) random-permutation multiplicative cascade model defined on a $d$-dimensional regular lattice. We start with a single cell of unit mass and partition it into $n=b^d$ child cells, where $b$ is the number of partitions in each dimension $d$. Next, we prepare a fixed set of weights $\{p_i\}_{i=1}^{n}$ with $\sum_i p_i=1$, shuffle them in random order, and assign one to each child cell. The mass of each child cell is then equal to the mass of the parent cell multiplied by the assigned weight. This process is repeated $L$ times, independently within each cell. The resulting field is defined on a lattice with $b^{L+1}$ cells on each edge, and we write $\rho$ for the cell mass normalized to unit mean. Hereinafter, we define the depth as $j = L + 1$, where $L$ is the total number of multiplication steps. Therefore, the level-$j$ field is defined on a lattice with $b^{j}$ cells on each side. Since each block is always assigned the same permutation of the multiset, the sum of the masses of a cell's child cells is exactly equal to the mass of the parent cell. In other words, this cascade enforces \emph{exact local mass conservation} at every node, which corresponds to the microcanonical constraint.

This sets the model apart from two similar cases. In the canonical Mandelbrot
cascade~\cite{Mandelbrot1974} the children carry independent, identically
distributed weights, and thus mass is conserved only on average. The conservative split
is different because the siblings share a fixed set of weights: if one is heavy the
others must compensate, and thus their log-weights are anticorrelated,
$\mathrm{Cov}(\ln w_a,\ln w_b)/\mathrm{Var}(\ln w)=-1/(n-1)$ ($n=b^d$;
weight-independent, $-1/3$ for $n=4$). For a canonical cascade these correlations
vanish, so the non-zero $-1/(n-1)$ is the unambiguous signature of \emph{exact}
conservation. This is  exactly what allows us to close the joint-exceedance probabilities of
the genus calculation. And unlike a lognormal
field~\cite{ColesJones1991,Xavier2016}, the cascade is \emph{not} a pointwise
transform of a Gaussian field: it carries genuine inter-scale phase structure
imprinted by the nested permutations, which is precisely what its topology will
reveal. Figure~\ref{fig:cascades}(a--c) displays the $\log_{10}\rho$ fields of three
cascades of increasing singularity width on a shared permutation skeleton: a
Sierpinski monofractal, a narrow multifractal, and a wide one, ranging from a
fractal set to a strongly intermittent web of voids, ridges, and peaks.

\subsection{The exact multifractal spectrum}
\label{sec:spectrum}

For a multinomial measure the multifractal spectrum has a closed-form.
The mass exponent, generalized dimensions and singularity spectrum are
\begin{align}
\tau(q) &= -\log_b\!\Big(\textstyle\sum_{i} p_i^{\,q}\Big),
&D_q &= \frac{\tau(q)}{q-1}, \label{eq:tau}\\
\alpha(q) &= \frac{\mathrm{d}\tau}{\mathrm{d}q},
&f(\alpha) &= q\,\alpha-\tau(q), \label{eq:falpha}
\end{align}
with $D_0=\log_b m$ ($m$ the number of nonzero weights, so $D_0=d$ when all $n$ are positive), $D_1$ the information dimension, and the correlation dimension
$D_2=-\log_b(\sum_i p_i^2)$. The width of the singularity spectrum,
\begin{equation}
\Da = \log_b\!\big(p_{\max}/p_{\min}\big),
\label{eq:width}
\end{equation}
(with $p_{\max}$ and $p_{\min}$ the largest and smallest \emph{nonzero} weights) is a single scalar that measures how multifractal the field is, and it serves
throughout as our non-Gaussianity dial. Equations~\eqref{eq:tau}--\eqref{eq:width}
are textbook multifractal theory~\cite{Mandelbrot1974,FrischParisi1985,Halsey1986};
we reproduce them only to fix notation. The Chhabra--Jensen direct
method~\cite{ChhabraJensen1989} recovers them to machine precision, because for the
multinomial measure $Z_q(b^{-n})=(\sum_i p_i^q)^n$ identically
[Figs.~\ref{fig:cascades}(d)--\ref{fig:cascades}(f)].

\section{The exact digital genus}
\label{sec:genus}

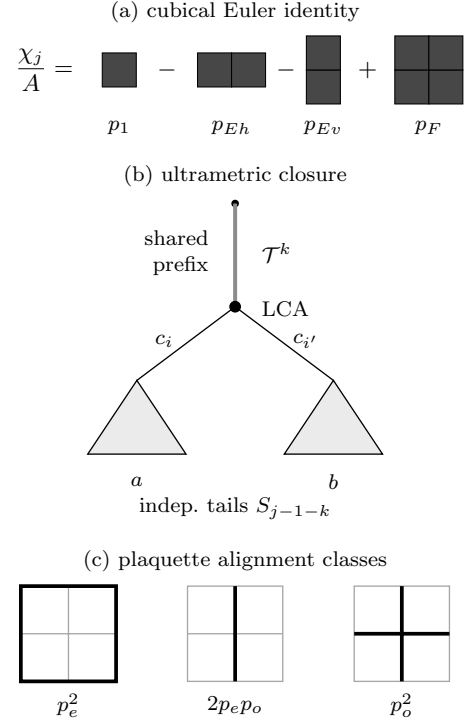
\begin{figure}
{\footnotesize (a) cubical Euler identity}\\[4pt]
\begin{tikzpicture}[x=0.30cm,y=0.30cm,line width=0.4pt, scale=1.50]
  \node[anchor=east] at (-0.6,1) {$\dfrac{\chi_j}{A}=$};
  \filldraw[fill=black!72] (0,0.5) rectangle (1,1.5);
  \node[font=\footnotesize] at (0.5,-0.7) {$p_1$};
  \node at (1.9,1) {$-$};
  \filldraw[fill=black!72] (2.8,0.5) rectangle (3.8,1.5);
  \filldraw[fill=black!72] (3.8,0.5) rectangle (4.8,1.5);
  \node[font=\footnotesize] at (3.8,-0.7) {$p_{Eh}$};
  \node at (5.4,1) {$-$};
  \filldraw[fill=black!72] (6.0,0) rectangle (7.0,1);
  \filldraw[fill=black!72] (6.0,1) rectangle (7.0,2);
  \node[font=\footnotesize] at (6.5,-0.7) {$p_{Ev}$};
  \node at (7.8,1) {$+$};
  \filldraw[fill=black!72] (8.6,0) rectangle (9.6,1);
  \filldraw[fill=black!72] (9.6,0) rectangle (10.6,1);
  \filldraw[fill=black!72] (8.6,1) rectangle (9.6,2);
  \filldraw[fill=black!72] (9.6,1) rectangle (10.6,2);
  \node[font=\footnotesize] at (9.6,-0.7) {$p_F$};
\end{tikzpicture}
\\[4pt]
{\footnotesize (b) ultrametric closure}\\[4pt]
\begin{tikzpicture}[line width=0.5pt,>=latex,font=\footnotesize, scale=1.30]
  \fill (0,0) circle (1.1pt);
  \draw[line width=1.5pt,black!45] (0,0) -- (0,-1.05);
  \node[anchor=east,align=right] at (-0.18,-0.5) {shared\\prefix};
  \node[anchor=west] at (0.18,-0.5) {$\mathcal{T}^{k}$};
  \fill (0,-1.05) circle (1.7pt);
  \node[anchor=west] at (0.18,-1.07) {LCA};
  \draw (0,-1.05) -- (-1.0,-1.8);
  \draw (0,-1.05) -- ( 1.0,-1.8);
  \node at (-0.72,-1.4) {$c_i$};
  \node at ( 0.72,-1.4) {$c_{i'}$};
  \filldraw[fill=black!8] (-1.0,-1.8) -- (-1.5,-2.55) -- (-0.5,-2.55) -- cycle;
  \filldraw[fill=black!8] ( 1.0,-1.8) -- ( 0.5,-2.55) -- ( 1.5,-2.55) -- cycle;
  \node at (-1.0,-2.83) {$a$};
  \node at ( 1.0,-2.83) {$b$};
  \node at (0,-3.1) {indep.\ tails $S_{j-1-k}$};
\end{tikzpicture}
\\[4pt]
{\footnotesize (c) plaquette alignment classes}\\[4pt]
\begin{tikzpicture}[x=0.42cm,y=0.42cm,line width=0.5pt,font=\footnotesize, scale=1.50]
  \draw[black!35] (1,0)--(1,2) (0,1)--(2,1);
  \draw[line width=1.4pt] (0,0) rectangle (2,2);
  \node at (1,-0.5) {$p_e^2$};
  \draw[black!35] (3.5,0) rectangle (5.5,2);
  \draw[black!35] (3.5,1)--(5.5,1);
  \draw[line width=1.4pt] (4.5,0)--(4.5,2);
  \node at (4.5,-0.5) {$2p_ep_o$};
  \draw[black!35] (7,0) rectangle (9,2);
  \draw[line width=1.4pt] (8,0)--(8,2) (7,1)--(9,1);
  \node at (8,-0.5) {$p_o^2$};
\end{tikzpicture}
\caption{The construction and its closure. \emph{(a)} The cubical Euler identity,
Eq.~\eqref{eq:cubchi}: the genus density is the ON-cell (vertex) probability $p_1$
minus the horizontal and vertical adjacent-pair (edge) probabilities $p_{Eh},p_{Ev}$
plus the $2\times2$ (face) probability $p_F$, the digital $V-E+F$ counted per unit
area. \emph{(b)} The ultrametric closure of the pair term, Eq.~\eqref{eq:pair}: two
adjacent cells $a,b$ share an ancestral prefix of $k$ levels (common mass
$\sim\mathcal{T}^{k}$), split at their lowest common ancestor (LCA) into two sibling
increments $c_i\neq c_{i'}$ drawn \emph{without replacement} (the functional
$\Psi_2$), then evolve down independent sub-cascades ($S_{j-1-k}$). \emph{(c)} The
three alignment classes of a $2\times2$ block on the dyadic grid, with their
finite-level weights, that drive the plaquette recursion Eq.~\eqref{eq:plaq}: both
corners even (all four cells share one parent, weight $p_e^2$, the $\Phi_4$ product);
mixed (two vertical pairs, weight $2p_ep_o$, the $G[\Psi_2]$ term); both odd (four
separate parents, weight $p_o^2$, the $F[\Psi_1]$ recursion).}
\label{fig:closure}
\end{figure}

\subsection{The cubical Euler identity}

Consider the level-$j$ excursion set: the cells whose  density exceeds a
threshold $\nu$. For a two-dimensional digital set the Euler characteristic is additive over $2\times2$ pixel
windows~\cite{OhserMucklich2000}, so the Euler density (genus per unit area) is
exactly
\begin{equation}
\frac{\chi_j(\nu)}{A} = p_1 - p_{Eh} - p_{Ev} + p_F ,
\label{eq:cubchi}
\end{equation}
where $p_1=\Pr(\rho>\nu)$ is the chance that a single cell is ON (the area
fraction), $p_{Eh}$ and $p_{Ev}$ are the chances that two horizontally or
vertically adjacent cells are both ON, and $p_F$ is the chance that all four cells
of a $2\times2$ block are ON. Here $\chi_j/A$ is the bulk (boundary-free) Euler density: $p_1$, $p_{Eh}$, $p_{Ev}$ and $p_F$ are per-cell, per-adjacent-pair and per-block exceedance probabilities, so Eq.~\eqref{eq:cubchi} is the $N\to\infty$ limit in which the pair and block counts per cell tend to one. On a finite $N\times N$ grid the raw counts carry boundary factors $(N-1)/N$ and $[(N-1)/N]^2$; we use these per-configuration densities consistently in both prediction and measurement, so the factors cancel. By the $x$--$y$ symmetry of the cascade
$p_{Eh}=p_{Ev}$. So the genus is fixed by just three joint exceedance
probabilities [Fig.~\ref{fig:closure}(a)] and the cascade's recursive structure closes each in closed form.

\begin{figure*}
\centering
\includegraphics[width=0.95\textwidth]{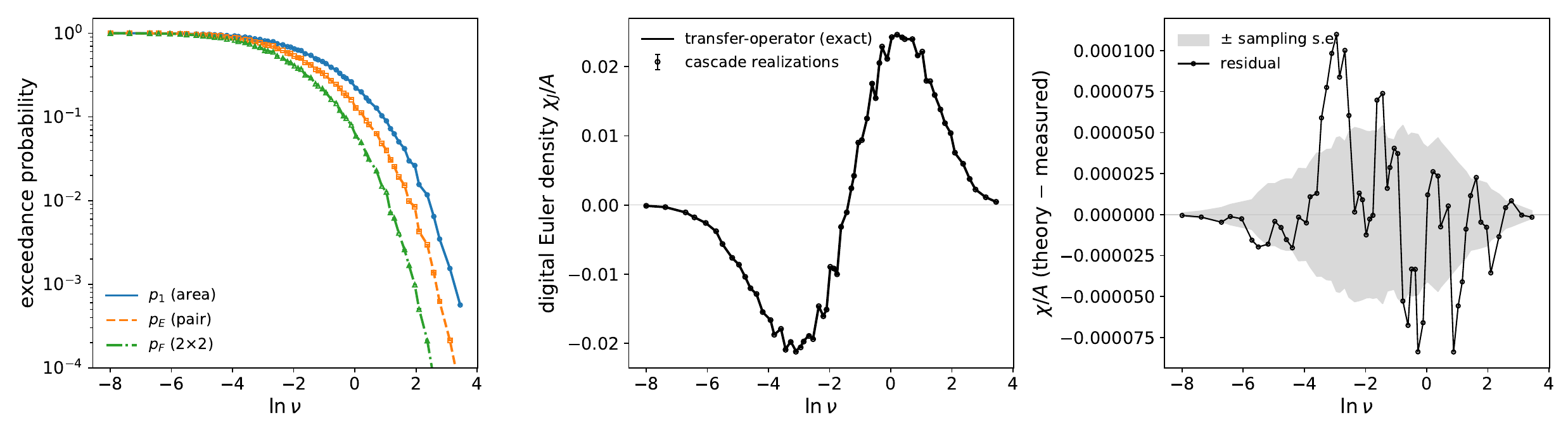}
\caption{The exact digital genus. The cubical Euler density $\chi_j/A$ of
Eq.~\eqref{eq:cubchi}, evaluated by the transfer operator (lines), versus direct
cascade realizations (points), for $b=2$ weights $[0.42,0.30,0.20,0.08]$ at
$j=8$. \emph{Left:} the one-, two-, and four-cell exceedance probabilities.
\emph{Centre:} the assembled genus curve (thresholds placed in the gaps between
the dominant atoms), agreeing to a maximum residual $\sim\!10^{-4}$ and not
relaxing to the Gaussian shape. \emph{Right:} the residual (theory $-$
measured) and the sampling-error band: a systematic discretization effect (the
measured genus is a step function of the atomic measure, the operator its
smooth envelope), which therefore does not shrink with the number of
realizations (Sec.~\ref{sec:validation}).}
\label{fig:digital}
\end{figure*}

\subsection{The transfer operator and its ultrametric closure}
\label{sec:closure}

We first need the one-cell statistics. The key simplification is that a logarithm
turns the product of weights into a sum: the log-density of a level-$j$ cell is a
sum of $j$ independent log-multipliers, $\ln\rho=\sum_{k=1}^{j}\ln(b^d W_k)$. The
distribution of such a sum is generated, one level at a time, by the transfer
operator
\begin{equation}
(\Tcal P)(x) = \frac1n\sum_{i=1}^{n} P\!\big(x-\ln(b^d p_i)\big),
\qquad n=b^d,
\label{eq:transfer}
\end{equation}
which simply shifts the current distribution by each possible increment and
averages. Starting from a spike at the origin, $P_j=\Tcal^{\,j}\delta_0$ is the
exact law of a level-$j$ cell, and the one-cell term is its tail,
$p_1(\nu)=\int_{\ln\nu}^{\infty}P_j$.

The joint two- and four-cell probabilities are where the tree earns its keep. Two
neighboring cells are not independent: they share a common ancestor
[Fig.~\ref{fig:closure}(b)]. Concretely,
they share an ancestral prefix of $k$ levels (a common mass drawn from $\Tcal^k$),
then split apart at a single branching, where they receive two \emph{distinct}
sibling weights drawn without replacement (the conservative split), and finally
evolve down independent sub-cascades. We can turn this picture directly into a
formula. Writing $\psi(x)=\mathbf{1}[x>\ln\nu]$ for the threshold test and $S_m$ for
the $m$-fold independent-tail smoothing, $(S_m\phi)(x)=\sum_{a}P_m(a)\,\phi(x+a)$
with the sum running over the atoms $a$ of $P_m$, the pair probability is
$p_{Eh}=G_j[\psi]$ with
\begin{equation}
G_j[\phi]=\sum_{k=0}^{j-1}\frac{2^k}{2^j-1}\,
\big\langle P_k,\;\Psi_2[\,S_{j-1-k}\phi\,]\big\rangle ,
\label{eq:pair}
\end{equation}
where the weight $2^k/(2^j-1)$ is the exact distribution of the common-ancestor
depth of an adjacent pair on an open line (most pairs branch apart near the bottom,
so deep ancestors dominate), and the conservative sibling-pair functional is
\begin{equation}
\Psi_2[\phi] = \frac{1}{n(n-1)}\!\sum_{i\neq i'}\!\phi(\cdot+c_i)\,\phi(\cdot+c_{i'}),
\qquad c_i=\ln(b^d p_i),
\label{eq:psi2}
\end{equation}
i.e.\ the two siblings carry a pair of increments drawn without replacement.

The four-cell block $p_F$ is handled by a recursion. How the four cells sit in the
tree depends on where the block falls relative to the dyadic grid lines [its
\emph{alignment class}; Fig.~\ref{fig:closure}(c)]. On an open grid of side $2^j$ a block corner lands on an
even index with probability $p_e=2^{j-1}/(2^j-1)$ and on an odd one with
$p_o=(2^{j-1}-1)/(2^j-1)$, giving
\begin{align}
F_j[\phi] = {}& p_e^2\,\big\langle P_{j-1},\,\Phi_4[\phi]\big\rangle
+ 2\,p_e p_o\, G_{j-1}\!\big[\Psi_2[\phi]\big] \notag\\
&{}+ p_o^2\, F_{j-1}\!\big[\Psi_1[\phi]\big],
\label{eq:plaq}
\end{align}
where $\Phi_4[\phi]=\prod_i\phi(\cdot+c_i)$ is the all-four-children product,
$\Psi_1[\phi]=\frac1n\sum_i\phi(\cdot+c_i)$ the single-child average, $G$ the pair
functional of Eq.~\eqref{eq:pair}, and $p_F=F_j[\psi]$. The weights
$(p_e^2,\,2p_ep_o,\,p_o^2)$ tend to the asymptotic dyadic fractions
$(\tfrac14,\tfrac12,\tfrac14)$ as $j\to\infty$, but we must keep their $O(1/2^j)$
finite-grid correction for the prediction to match a finite realization exactly.

This construction specializes the tree-sum cumulant machinery of Greiner
\emph{et al.}~\cite{GreinerEggersLipa1998,Greiner1998} from correlation cumulants
to the topological exceedance functionals of Eq.~\eqref{eq:cubchi}. The
without-replacement split in Eq.~\eqref{eq:psi2} is the geometric counterpart of the
microcanonical anticorrelation $-1/(n-1)$ of Sec.~\ref{sec:model}. Collecting the
pieces gives our central result.

\begin{result}[Exact digital genus]
\label{res:genus}
For a two-dimensional conservative binary ($b=2$, $n=4$) random-permutation
cascade with weight multiset $\{p_i\}$, the expected digital Euler density of the
level-$j$ excursion set at any threshold $\nu$ is given exactly by
Eq.~\eqref{eq:cubchi}, with $p_1$ the $\Tcal^j$ tail and $p_{Eh}=p_{Ev}$, $p_F$
the finite functionals of Eqs.~\eqref{eq:pair}--\eqref{eq:plaq}. The result
involves no smoothing and no Monte Carlo, and is evaluated in time polynomial in
$j$: the transfer-operator tails and the multinomial atoms cost $O(j^{\,n-1})$
each, and the plaquette recursion has depth $j$. The closure itself is not
specific to $b=2$ or $d=2$: for a general base the ancestral-depth weights
become $b^k(b-1)/(b^{\,j}-1)$, and the three-dimensional case follows the same
route (Sec.~\ref{sec:discussion}).
\end{result}

\section{Validation and the residual}
\label{sec:validation}

We now check the prediction against direct cascade simulations.
Figure~\ref{fig:digital} compares Eq.~\eqref{eq:cubchi}, evaluated by the transfer
operator, with the genus measured from realizations. The three joint exceedance probabilities and the assembled genus curve agree with a maximum residual of $\sim\!10^{-4}$ in $\chi_j/A$, and $p_1$ is exact to machine precision. The genus is clearly non-Gaussian: it does not relax to
the antisymmetric Tomita shape $\nu\,e^{-\nu^2/2}$, but keeps the asymmetric
void/peak balance of the intermittent measure.

This small residual is related to what we mean by the term \emph{exact}. The bare measure is atomic under the dyadic measure. That is, a level-$j$ cell can take on only one of a finite number of multinomial density values. Therefore, the measured genus $\chi_j(\nu)$ is actually a \emph{step function} with respect to the threshold, whereas the transfer operator returns a smoothed \emph{envelope} of this step function. The area fraction $p_1$ is an exception, as it uses the same strict inequality to threshold exact atoms, resulting in a residual of exactly zero. The pair and plaquette terms $p_{Eh,v}$ and $p_F$, evaluated on the lattice, differ from the measured values only where a threshold lands within the lattice resolution of a heavy atom; even then, the difference is only half the size of the local jump. This is purely a discretization effect. Since this effect is reproduced in independent realizations, it is not sampling noise; furthermore, since it does not appear in $p_1$, it is not a model error either. When the threshold is placed at the spacing between the main atoms, as in this study, the comparison takes place in the flat parts of the step function, and the residual remains at the level of $\sim 10^{-4}$. Furthermore, two finite-level effects contributing at the same order must be included. The first is the open-line ancestor depth distribution shown in Eq.~\eqref{eq:pair} (where applying periodic wrapping adds artificial ``seam'' pairs with depth zero), and the second is the exact alignment fractions shown in Eq.~\eqref{eq:plaq}.

\begin{figure}
\includegraphics[width=0.45\textwidth]{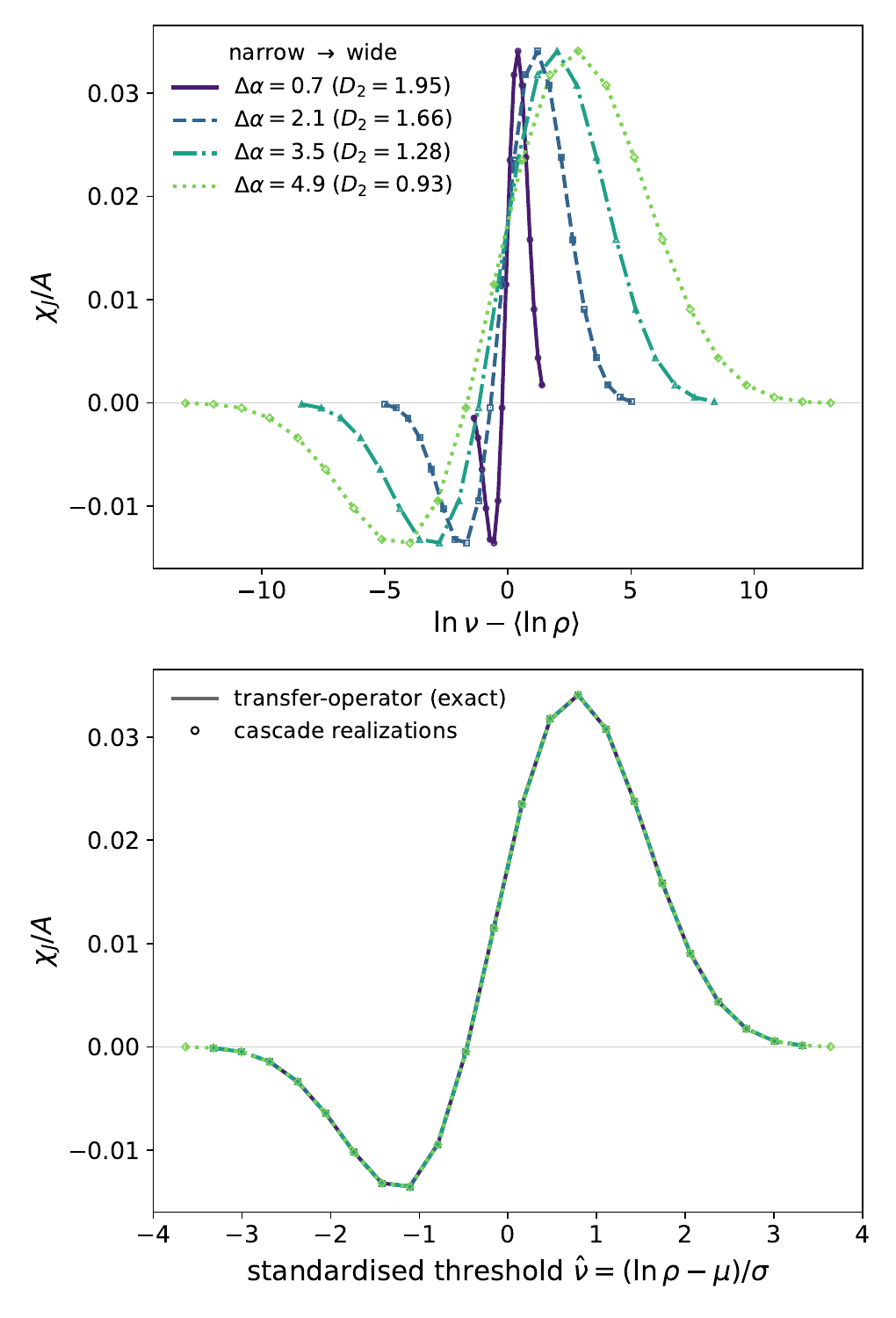}
\caption{The exact digital genus for a geometric weight
family $p_i\propto r^{\,i}$ swept from near-monofractal ($\Da=0.7$) to strongly
intermittent ($\Da=4.9$); theory (lines) versus realizations (points).
\emph{Top:} centered on the log-density mean, the genus keeps its shape and
amplitude but stretches in $\ln\nu$ as the width grows. \emph{Bottom:} under the
standardized threshold $\hat\nu$ the curves collapse onto one universal shape:
the digital genus is self-similar under the width dial, with a width-invariant
topological amplitude.}
\label{fig:width}
\end{figure}

\section{Scaling and self-similarity}
\label{sec:scaling}

\subsection{Tie to the multifractal spectrum}

The shape of the genus curve is not arbitrary: its threshold scaling is set by the
spectrum. Writing $\theta=\ln\nu$, the rare-density tail of a cell is governed by a large-deviation rate that is the Legendre transform of $\tau(q)$. At leading (large-deviation) order the logarithmic slope of the relevant tail probability approaches
\begin{equation}
\frac{\mathrm{d}\ln p_1}{\mathrm{d}\theta} \;\xrightarrow[\,j\to\infty\,]{}\; -q, \qquad \alpha(q)=d-\frac{\theta}{j\ln b}\quad(q>0),
\label{eq:tailslope}
\end{equation}
so the slope at a given threshold reads off the moment order $q$ whose singularity $\alpha(q)=\tau'(q)$ sits at that density; at finite $j$ the multinomial-tail prefactor adds $O(1/j)$ corrections. The upper tail $p_1=\Pr(\rho>\nu)$ (the $q>0$ branch) governs the peak lobe of the genus, and the void lobe reads the corresponding lower-tail rate $\Pr(\rho<\nu)$ ($q<0$).

\subsection{Self-similarity under the width dial}

For a \emph{geometric} sequence of weights, where $p_i\propto r^{\,i}$, an interesting structural result follows. In such a sequence, the logarithmic increments at each level
$c_i=\ln(b^d p_i)=\mathrm{const}+i\ln r$
are equally spaced; therefore, adjusting the width parameter (i.e., varying $r$) has the sole effect of stretching and shifting the $\ln\nu$ axis. Consequently, the genus exhibits exact self-similarity. That is, its amplitude (the maximum and minimum values of $\chi_j/A$) is a fixed combinatorial constant of the $n$-permutation tree, whereas its width increases linearly with $\Da$. To demonstrate this explicitly, we can normalize the threshold using the mean and standard deviation of the log density. That is,
$\hat\nu=(\ln\nu-\mu)/\sigma$,
where $\mu=j\langle c\rangle$ and
$\sigma^2=j\,\mathrm{Var}(c)$. When plotted against $\hat\nu$, the genus curves for all widths completely overlap (Fig.~\ref{fig:width}). The transfer operator reproduces this collapse with an accuracy of approximately $\sim 10^{-4}$ in the range $\Da=0.7$--$4.9$. This collapse holds exactly only for the equal-spacing family. In the case of a general multiset, the width does not simply stretch the curve but alters the \emph{shape} of the genus itself.

\section{Discussion}
\label{sec:discussion}

We have shown that the topology of a conservative multiplicative cascade can be exactly predicted at its native scale. The digital Euler characteristic of the
level-$j$ excursion set is a finite, closed-form functional of a transfer operator
on the tree, closed by the ultrametric ancestral structure, the conservative
sibling split, and the exact plaquette alignment combinatorics. We show that it matches
simulations down to the discretization floor. Thus, the multiplicative cascade, along with the Rayleigh--L\'evy flight, takes its place as a controllable and fully specified non-Gaussian benchmark model.
The two approaches are complementary. For additive flight, the Euler characteristics are currently known only from mean-field results~\cite{Bernardo2026}, whereas for the multiplicative cascade, an exact solution exists.

Several extensions follow immediately. The closure property in Sec.~\ref{sec:closure} does not depend on dimension. In three dimensions, the Euler density of a cube is additive with respect to $2\times2\times2$ windows, and each of the eight joint exceedance terms is closed by the same ancestral-prefix factorization. Therefore, the three-dimensional genus can also be exactly solved in the same manner. The same tree structure closes the density correlation function hierarchy at all orders. That is, a connected $n$-point function is represented as the sum of rooted tree topologies with $n$ leaves, and each term contains a corresponding joint without replacement. We leave these extensions, and the cosmological application that motivates the model (the cascade as a fast, tunable surrogate for the non-Gaussian morphology of the cosmic density field, to be compared with $N$-body halo fields), to future work. Smoothing destroys the multiplicative tree structure (for a smoothing kernel $K$ and a log density $\Gamma=\ln\rho$, $K*e^{\Gamma}\neq e^{K*\Gamma}$). Therefore, even when an exact value for skewness is given, the genus of the smoothed field does not have a closed-form solution. The exactly analyzable digital genus derived in this study provides a closed-form anchor for interpreting the morphology of empirically measured smoothed fields.

\begin{acknowledgments}

The cascade generator and topology code are available at
\url{https://github.com/csabiu/multiplicative_cascade}.

We thank Luis Teodoro, Christophe Pichon and Stephen Appleby for stimulating discussions regarding this work. 

C.G.S. acknowledges support from the Basic Science Research Program
(2018R1A6A1A06024977) through Korea's NRF funded by the Ministry of Education.

\end{acknowledgments}

\bibliographystyle{apsrev4-2}
\bibliography{refs}

\end{document}